\documentclass[english]{article}
\usepackage{mathptmx}
\usepackage[T1]{fontenc}
\usepackage[latin9]{inputenc}
\usepackage[a4paper]{geometry}
\usepackage{textcomp}
\usepackage{graphicx}
\usepackage{setspace}
\makeatletter

\DeclareRobustCommand{\greektext}{%
  \fontencoding{LGR}\selectfont\def\encodingdefault{LGR}}
\DeclareRobustCommand{\textgreek}[1]{\leavevmode{%
  \IfFileExists{grtm10.tfm}{}{\fontfamily{cmr}}\greektext #1}}
\DeclareFontEncoding{LGR}{}{}
\DeclareTextSymbol{\~}{LGR}{126}
\providecommand{\tabularnewline}{\\}

\newcommand{\lyxaddress}[1]{
\par {\raggedright #1
\vspace{1.4em}
\noindent\par}
}

\makeatother

\usepackage{babel}
\begin{document}

\title{Simulation of High Conversion Efficiency and Open-circuit Voltages
Of $\alpha$-si/poly-silicon Solar Cell}

\author{AQing Chen$^{1}$, QingYi Shao$^{2}$%
\thanks{Corresponding author (email: qyshao@163.com)Tel: +86 013826089736%
}}

\maketitle

\lyxaddress{\begin{center}
$^{1}$School of New Energy Engineering, Leshan Vocational \& Technical
College, Leshan 614000, China $^{2}$Laboratory of Quantum Information
Technology, School of Physics and Telecommunication Engineering, South
China Normal University, Guangzhou 510006, China
\par\end{center}}
\begin{abstract}
The P$^{+}$ $\alpha$-Si /N$^{+}$polycrystalline solar cell is molded
using the AMPS-1D device simulator to explore the new high efficiency
thin film poly-silicon solar cell. In order to analyze the characteristics
of this device, and the thickness of N$^{+}$ poly-silicon, we consider
the impurity concentration in the N$^{+}$ poly-silicon layer and
the work function of transparent conductive oxide (TCO) in front contact
in the calculation. The thickness of N$^{+}$ poly-silicon had little
impact on the device when the thickness varied from 20\textgreek{m}m
to 300\textgreek{m}m. The effects of impurity concentration in polycrystalline
are analyzed. The conclusion is drawn that the open-circuit voltages
(Voc) of P$^{+}$ \textgreek{a}-Si /N$^{+}$ polycrystalline solar
cell is very high reaching 752 mV, and the conversion efficiency reaches
9.44\%. Therefore, based on the above optimum parameters the study
on the device formed by P$^{+}$ $\alpha$-Si/N$^{+}$ poly-silicon
is significant in exploring the high efficiency poly-silicon solar
cell.

\textbf{keywords: }P$^{+}$ $\alpha$-Si/N$^{+}$ poly-silicon, Solar
cell, Photovoltaics, Thin-film polycrystalline-silicon

\end{abstract}

\section{Introduction}

Thin-film polycrystalline-silicon solar cells have offered a promising
alternative to standard silicon solar cells. Up to now, however, obtained
efficiencies are too low to cut the cost in photovoltaics. Poly-crystalline
silicon solar cell modules currently represent between 80\% and 90\%
of the PV world market resulting from the stability, robustness and
reliability of this kind of solar cells compared to those of emerging
technologies. Thin-film polycrystalline-silicon solar cells are considered
to be one of the most promising alternatives to bulk silicon solar
cells. Thin films decrease the cost of silicon wafers significantly
so that they account for about half of the total cost of standard
silicon solar modules. There are many methods for preparing the thin-film
polycrystalline-silicon. Both B-doped and n-type polycrystalline silicon
thin films were prepared by rapid thermal chemical vapor deposition
\cite{ai_electrical_2006} and chemical vapor deposition (CVD) \cite{tuezuen_properties_2010},
respectively. In light of these merits, the thin film polycrystalline-silicon
solar cells have been studied through many methods. Thin-film solar
cells which are fabricated by polycrystalline silicon (poly-Si) films
prepared by flash lamp annealing have been fabricated by Yohei Endo
et al \cite{endo_thin-film_2010}. They had the open circuit voltage
(Voc) of 0.21 V and fill factor (FF) of 0.404. Thin-film polycrystalline
silicon solar cells on ceramic studied by L. Carnel \cite{carnel_thin-film_2006}
had the maximum efficiency of 5.3\% and Open-circuit voltages (Voc)
of 520 mV. S. Gall et al \cite{gall_polycrystalline_2009} prepared
the polycrystalline silicon thin-film solar cells with high conversion
efficiencies on glass using innovative approaches. The studies on
polycrystalline silicon thin-film solar cells have obviously focused
on photovoltaic industry. Exploring the high conversion efficiency
and large open-circuit voltages (Voc) is of importance for the application
of polycrystalline silicon thin-film solar cells. However, the high
open-circuit voltages and high conversion efficiency can not be achieved
simultaneously for thin-film polycrystalline silicon solar cells.
For this reason, we have designed a special polycrystalline silicon
solar cell to achieve the required criterion. 

The thin film amorphous silicon which is also applied in solar cell
can be obtained at a low temperature by hot-wire chemical vapor deposition
\cite{villar_amorphous_2009} or plasma enhanced chemical vapor deposition
(PECVD) \cite{swatowska_amorphous_2008}. We assumed that both Voc
and conversion efficiency were improved if the P$^{+}$ thin film
amorphous silicon was deposited on the N$^{+}$ polycrystalline silicon
forming a solar cell. In order to prove our assumption, the frame
of P$^{+}$ thin film amorphous silicon/ N$^{+}$ poly-silicon was
simulated using the AMPS-1D software package \cite{zhu_applications_1999}
which is a very general computer simulation code for analyzing and
designing two terminal structures. It can handle devices such as p\textendash{}n
and p\textendash{}i\textendash{}n homo- and hetero-junctions, p\textendash{}i\textendash{}p
and n\textendash{}i\textendash{}n structures, multi-junction and Schottky
barrier devices. These devices may have poly-crystalline, amorphous
or single crystal layers or their combinations. Device operation may
be undertaken in dark or under light, and hence it is possible to
simulate the behavior of solar cells and photodiodes. This program
has been widely used to study device \cite{hernandez-como_simulation_2010},
\cite{walsh_new_2007}, \cite{kabir_effect_2010}. By simulating P$^{+}$
\textgreek{a}-Si/N$^{+}$ poly-silicon cell, the maximum Voc of 0.752
V and conversion efficiency of 9.44\%, respectively, were obtained
on the condition of the optimum doping concentration and proper thickness
of N$^{+}$ polycrystalline silicon. Those values are promising in
photovoltaics industry for polycrystalline silicon solar cell. On
the other hand, the thin film amorphous silicon and polycrystalline
silicon were prepared with a high deposition rate by pulsed plasma
and Hot-Wire CVD techniques \cite{madan_high_1998} and hot wire cell
method \cite{ichikawa_high-rate_2001} (deposition rates of 1.2 nm/s).
That means that the solar cell formed by P$^{+}$ thin film amorphous
silicon and N$^{+}$ polycrystalline silicon can be fabricated and
that the cost will be less than other cells. Therefore, our studies
on the solar cell formed by P$^{+}$ $\alpha$-Si/N$^{+}$ poly-silicon
are significant for the application of polycrystalline silicon in
photovoltaics industry. 

This paper contains four sections. In section II, the structure and
the parameters of P$^{+}$ $\alpha$-Si/N$^{+}$ poly-silicon solar
cell were set. Section III analyzes the calculation results of the
model to characterize the P$^{+}$ $\alpha$-Si/N$^{+}$ poly-silicon
solar cell.

\section{P+ $\alpha$-Si/N+ poly-silicon solar cell model}

To simulate the P$^{+}$ $\alpha$-Si/N$^{+}$ poly-silicon solar
cell structure, we build the model like in figure 1. The thin film
amorphous silicon covers the polycrystalline silicon with the thickness
of 5-20nm. The thickness of polycrystalline silicon is on the order
of \textgreek{m}m. The P$^{+}$ thin film amorphous silicon is considered
as the emitter layer and the N$^{+}$ polycrystalline silicon as the
absorption layer. The light of the sun travels from the top to the
bottom. The first layer viewed as front contact is transparent conductive
oxide (TCO) consisting of the Indium tin oxide (ITO) which is the
most commonly used material in photovoltaic (PV) devices and light
emitting diodes (LED) as high work function transparent electrode.
AMPS-1D has two pictures, namely, the Lifetime and DOS pictures. In
the DOS picture, the details of recombination traffics, trapping and
the charge state of the defects (and the effects of this charge on
the electric field variation across a structure) are fully explained.
The DOS picture, hence, was employed to simulate the P$^{+}$ $\alpha$-Si/N$^{+}$
poly-silicon solar cell structure. All the parameters are listed in
Table 1 and the absorption coefficient for the poly-Si is obtained
from a research paper \cite{rajkanan_absorption_1979}. For P$^{+}$
$\alpha$-Si:H, the data is obtained from the compendium of parameters
of AMPS-1D, and the absorption coefficient for P$^{+}$ $\alpha$-Si:H
is the same as for intrinsic \textgreek{a}-Si:H \cite{searle_properties_1998}.
For the simulation under illumination we used the AM 1.5 spectrum
normalized to 100 mW/cm2.
\begin{figure}
\begin{centering}
\includegraphics{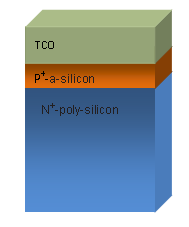}
\par\end{centering}

\caption{Structure of the P+ \textgreek{a}-Si/N+ poly-silicon solar cell}

\end{figure}

\section{Results and discussion}

In order to study the P$^{+}$ $\alpha$-Si/N$^{+}$ poly-silicon
solar cell well, we first consider the work function of TCO which
plays a great role in silicon hetero-junction solar cell performance
\cite{centurioni_role_2003}. ITO has a low work function compared
to the p type silicon material, usually varying from 4.35 eV to 5.1
eV. Figure 2 shows the various values of conversion efficiency at
different work functions of ITO. It is apparent that the obtained
efficiency increases with the work function of ITO when the value
of the work function is smaller than 4.95 eV. But the conversion efficiency
is independent of the work function of ITO when the value of the work
function is larger than 4.95 eV. So the work function should be controlled
above 4.95 eV when the P$^{+}$ $\alpha$-Si/N$^{+}$ poly-silicon
solar cell well is fabricated, which is very important for preparing
high efficiency solar cell.
\begin{table}
\caption{Parameters set for the simulation of the P+ \textgreek{a}-Si/N+ poly-silicon
solar cell structure with AMPS-1D}

\medskip{}

\centering{}%
\begin{tabular}{ccc}
\hline 
Parameter and units e, h for electrons and holes respectively  & P+ \textgreek{a}-Si & N+ poly-silicon\tabularnewline
\hline 
Thickness (nm)  & 5  & var. \tabularnewline
Electron affinity (eV) & 3.8  & 4.05 \tabularnewline
Band gap (eV)  & 1.72  & 1.12 \tabularnewline
Effective conduction band density (cm$^{-3}$)  & 2.5\texttimes{}10$^{20}$ & 2.8\texttimes{}10$^{19}$ \tabularnewline
Effective valence band density (cm$^{-3}$)  & 2.5\texttimes{}10$^{20}$ & 1.04\texttimes{}10$^{19}$ \tabularnewline
Electron mobility (cm$^{-2}$2 V$^{-1}$s$^{-1}$) & 10  & 750 \tabularnewline
Hole mobility (cm$^{2}$ V$^{-1}$s$^{-1}$)  & 1  & 250 \tabularnewline
Doping concentration of acceptors (cm$^{-3}$)  & 0  & var. \tabularnewline
Doping concentration of donors (cm$^{-3}$) & 1.0\texttimes{}10$^{20}$ & 0\tabularnewline
Band tail density of states (cm$^{-3}$eV$^{-1}$)  & 2.0\texttimes{}10$^{21}$  & 1.0\texttimes{}10$^{16}$\tabularnewline
Characteristic energy (eV)  & 0.06, 0.03  & 0.01,0.01 \tabularnewline
donors, acceptors Capture cross section for donor states, e, h, (cm$^{2}$)  & 1\texttimes{}10$^{-15}$, 1\texttimes{}10$^{-17}$  & 1.0\texttimes{}10$^{-14}$, 1.0\texttimes{}10$^{-16}$ \tabularnewline
Capture cross section for acceptor states, e, h, (cm$^{2}$)  & 1\texttimes{}10$^{-17}$, 1\texttimes{}10$^{-15}$  & 1.0\texttimes{}10$^{-16}$, 1.0\texttimes{}10$^{-14}$\tabularnewline
Gaussian density of states (cm$^{-3}$) & 8\texttimes{}10$^{17}$, 8\texttimes{}10$^{20}$  & \tabularnewline
Gaussian peak energy (eV) & 1.22, 0.70  & \tabularnewline
donors, acceptors Standard deviation (eV)  & 0.23  & \tabularnewline
Capture cross section for donor states, e, h, (cm$^{2}$)  & 1\texttimes{}10$^{-14}$, 1\texttimes{}10$^{-15}$  & \tabularnewline
Capture cross section for acceptor states, e, h, (cm$^{2}$) & 1\texttimes{}10$^{-15}$, 1\texttimes{}10$^{-14}$ & \tabularnewline
\hline 
\end{tabular}
\end{table}
\begin{figure}
\begin{centering}
\includegraphics[scale=0.3]{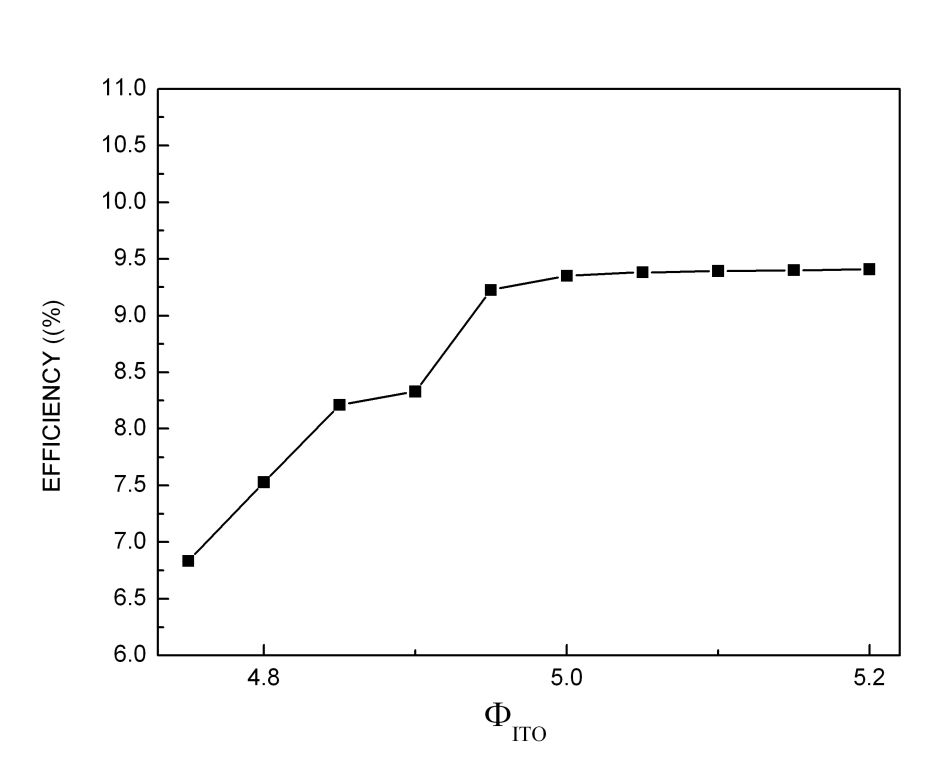}
\par\end{centering}

\caption{The conversion efficiency as a function of the work function of ITO
($\Phi_{ITO}$)}

\end{figure}

In order to find out the best parameters of this P$^{+}$ $\alpha$-Si/N$^{+}$
poly-silicon solar cell, we calculate the thickness of N$^{+}$ poly-silicon
and display its effects on this cell in Figure 3. It is easy to conclude
that the N$^{+}$ poly-silicon layer should be thick enough to absorb
light. If the layer is thin, the sun light can not be absorbed well,
leading to a low efficiency. On the other hand, this layer can not
be too thick. If it\textquoteright{}s too thick, the carries can not
arrive at the back contact owing to recombination. The best thickness
of the N$^{+}$ poly-silicon layer is the value of 20-300 $\mu$m.
\begin{figure}
\begin{centering}
\includegraphics[scale=0.5]{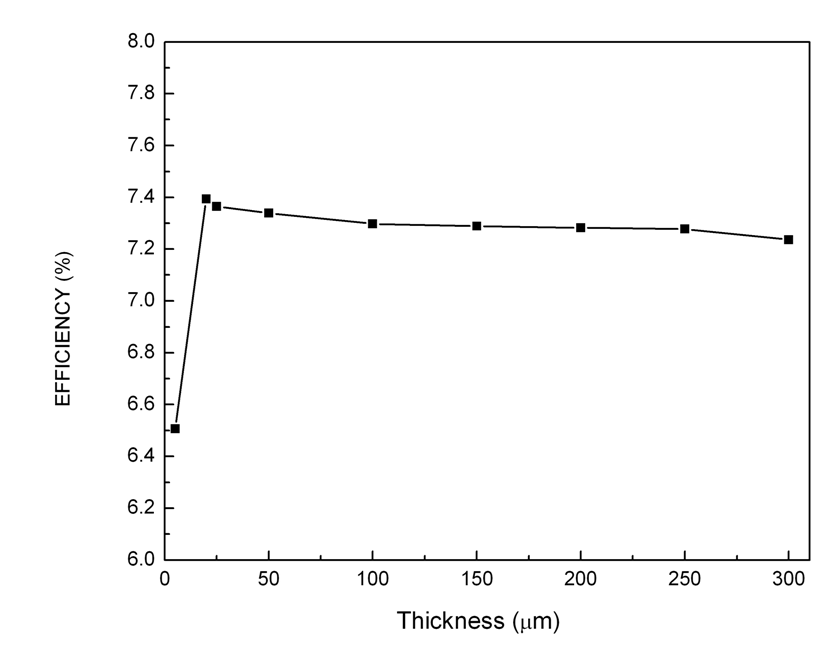}
\par\end{centering}

\caption{The conversion efficiency for different thickness of the N+ poly-silicon
layer}
\end{figure}

Figure 4 shows the effects of N$^{+}$ poly-silicon layer doping.
It can be seen that the conversation efficiency, fill factor and Voc
increase with the doping concentration ($N_{D}$) when $N_{D}<10^{19}$
cm$^{-3}$. But both efficiency and fill factor decrease when $N_{D}>10^{19}$
cm$^{-3}$. So the conversion efficiency gets up to the maximum value
of 9.44\% at $N_{D}=10^{19}$ cm$^{-3}$. The fill factor also has
the maximum value of 0.854 at $N_{D}=10^{19}$ cm$^{-3}$. What\textquoteright{}s
important is that the Voc can reach the maximum value of 0.752 V at
$N_{D}=10^{19}$ cm$^{-3}$. Although this value is not greater than
that of some poly-silicon solar cell which has high efficiency of
14.9\% \cite{elgamel_high-efficiency_1998}, the values of both fill
factor and Voc are greater than those of other poly-silicon solar
cells, which are usually about 0.5 V \cite{carnel_thin-film_2006}
for Voc and 0.6 V \cite{li_thin_2006} for fill factor.
\begin{figure}
\begin{centering}
\includegraphics[scale=0.1]{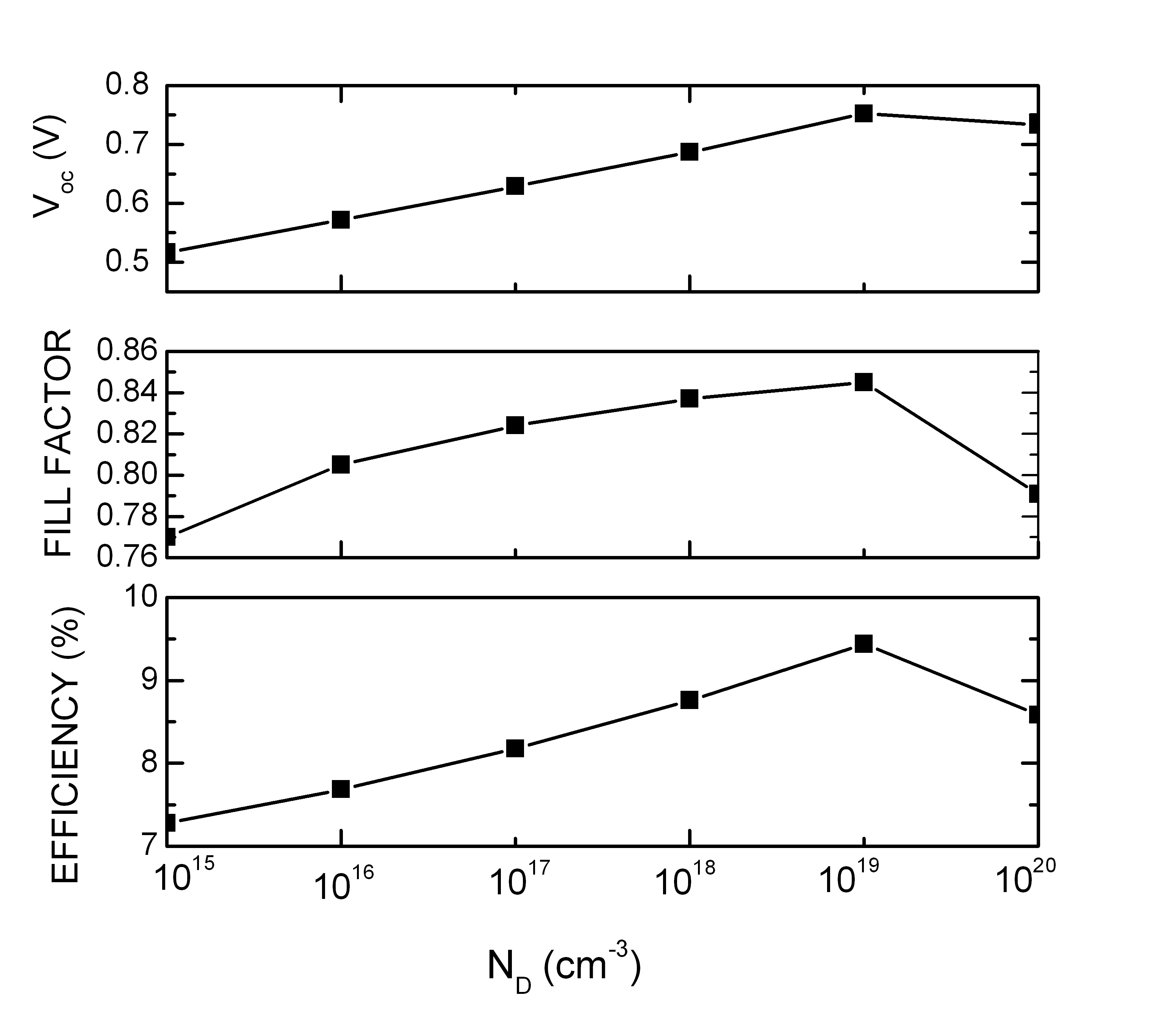}
\par\end{centering}

\caption{Values of efficiency, fill factor and Voc, respectively, are as a
function of doping concentration of N$^{+}$ poly-silicon layer}

\end{figure}
\begin{figure}
\begin{centering}
\includegraphics[scale=0.1]{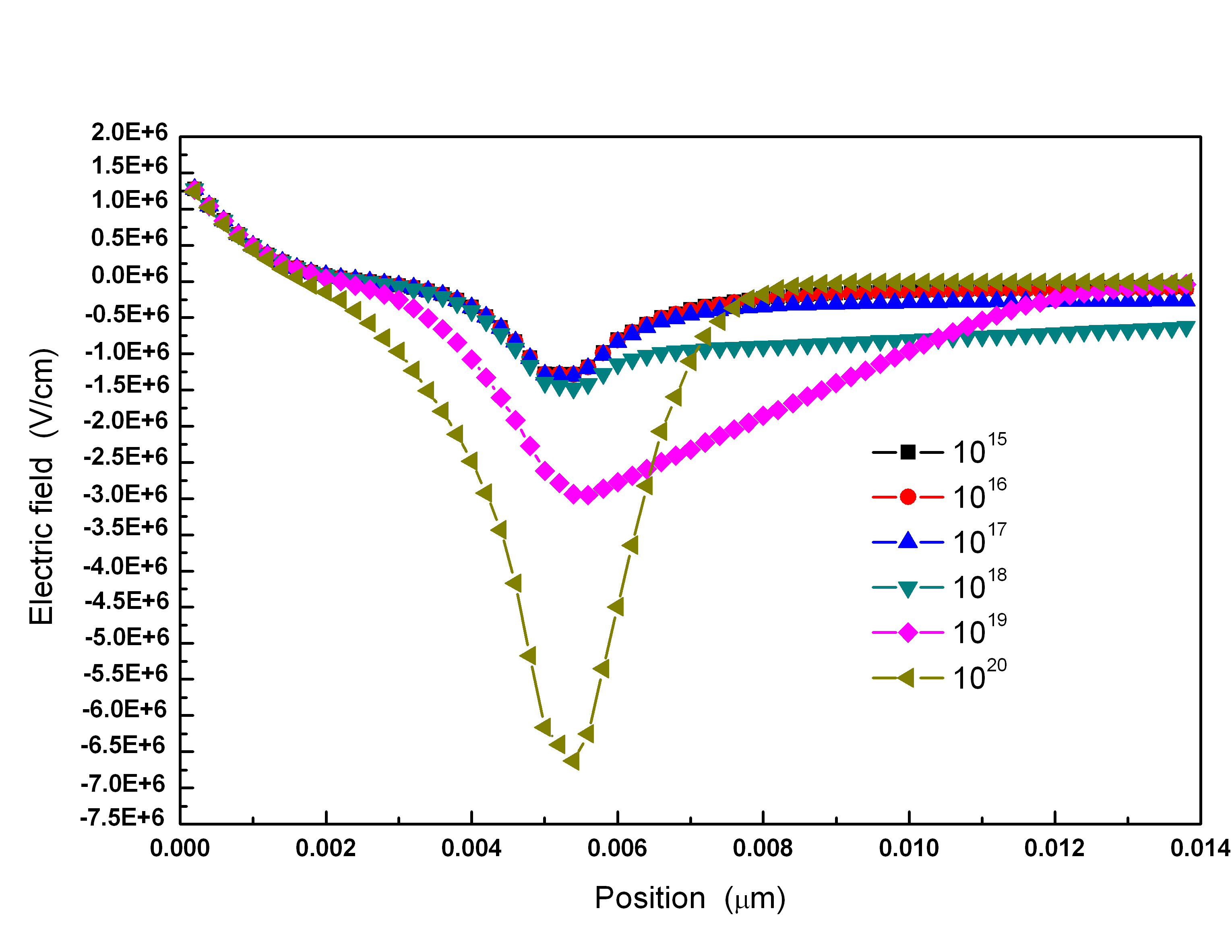}
\par\end{centering}

\caption{The distribution of the electric field with different doping concentrations
of the N$^{+}$ poly-silicon layer (N$_{n}$) as a function of position.}
\end{figure}
\begin{figure}
\begin{centering}
\includegraphics[scale=0.1]{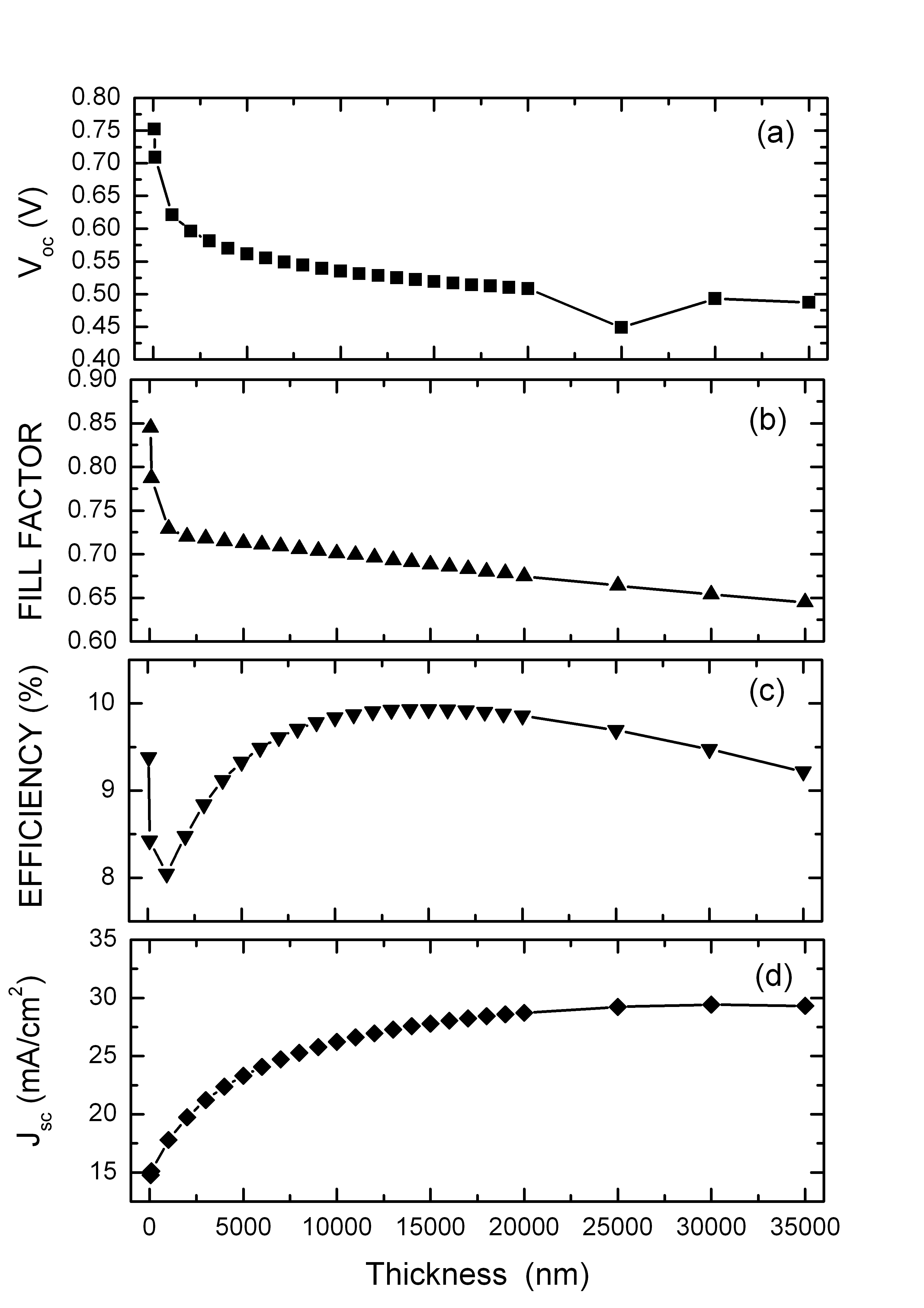}
\par\end{centering}

\caption{(a)-(d) illustrate the Voc (a), fill factor (b), conversation efficiency
(c) and short circuit current (Jsc) (d) of the P$^{+}$ $\alpha$-Si/N$^{+}$
poly-silicon solar cell with the different thickness of the intrinsic
layer, respectively.}
\end{figure}

In order to account for the effects of N$^{+}$ poly-silicon layer
doping on the P$^{+}$ $\alpha$-Si/N+ poly-silicon solar cell, the
distribution of electric field is taken into consideration, shown
in Figure 5. The electric field works as a function of the position.
When the concentration of the N$^{+}$ poly-silicon layer varies from
3.0\texttimes{}10$^{15}$ cm$^{-3}$ to 3.0\texttimes{}10$^{19}$
cm$^{-3}$, the electric field rises gradually and the area covered
by the electric field decreases little. That means the holes produced
in a wider area of the absorption layer can be transported through
the space charge zone. Although the electrical field increases sharply,
the area of the absorption layer becomes narrow, resulting in fewer
holes for transport once the concentration reaches 3.0\texttimes{}10$^{20}$
cm$^{-3}$.The above analysis indicates the best value of concentration
of N$^{+}$ poly-silicon layer is 3.0\texttimes{}10$^{19}$19 cm$^{-3}$
for high conversion efficiency. 

Because the intrinsic layer in a solar cell is to absorb sun light,
we allow changing the thickness to attain the effects of the intrinsic
layer on the P$^{+}$ $\alpha$-Si/N$^{+}$ poly-silicon solar cell.
Figures 6 (a)-(d) illustrate the Voc, fill factor, conversation efficiency
and short circuit current (Jsc) of the P$^{+}$ $\alpha$-Si/N$^{+}$
poly-silicon solar cell with the different thickness of the intrinsic
layer, respectively. Both open-circuit voltage and fill factor decrease
as the thickness of the intrinsic layer is inserted between P$^{+}$
$\alpha$-Si and N$^{+}$ poly-Si, as shown in Figures 6(a) and (b).
Why the intrinsic layer weakens the potential difference between P$^{+}$
$\alpha$-Si and N$^{+}$ poly-Si is explained, which leads to a smaller
value of open-circuit voltage. The thicker the intrinsic layer is,
the smaller the opencircuit voltage becomes. Besides, the whole resistance
of the cell increases with the thickness of the intrinsic layer, which
has a great impact on the fill factor. It is well known that when
an intrinsic poly-silicon layer is inserted between the p layer and
n layer, the conversation efficiency can be improved dramatically.
However, the conversation efficiency is not much improved for the
P$^{+}$ $\alpha$-Si/N$^{+}$ poly-silicon. Figure 6(c) shows the
relation between the conversation efficiency and the thickness of
the intrinsic layer. When the intrinsic layer is thin (about 1000
nm), the efficiency decreases sharply with the value of about 8\%.
While the thickness varies from 1000 nm to 15000 nm, the efficiency
increases gradually and reaches up to about 10\%. But the efficiency
will decrease gradually with the thickness of the intrinsic layer
when it is larger than 15000 nm. An in-depth discussion of the relationship
between the conversation efficiency and the thickness is made to find
out the reason.

There are many defects in P$^{+}$ $\alpha$-Si, intrinsic layer and
N$^{+}$ poly-Si resulting in a short lifetime of the carriers. Carriers
collection mainly depends on the in-build electric field when the
intrinsic layer is thin. But then the intrinsic layer can weaken the
electric field leading to a low efficiency. When the intrinsic layer
gets thicker, there are many photogenerated carriers. The carriers
are collected by the carriers diffusion, which will lead to a higher
conversation efficiency and larger short circuit current. Although
the carriers diffusion can be enhanced, the in-build electric field
is weakened by the thick intrinsic layer. Considering the two factors,
the conclusion can be drawn that the efficiency can not be improved
sharply when the intrinsic layer with proper thickness is inserted
between the P$^{+}$ $\alpha$-Si and N$^{+}$ poly-Si layer. The
above discussions lead to the important result that the intrinsic
layer can improve efficiency more or less but decrease the open circuit
voltage and fill factor.

\section{Conclusion}

This paper considers the three main effect factors responsible for
the outstanding characteristics of such a solar cell. They are the
work function of TCO, the thickness of N$^{+}$ the poly-silicon layer
and the doping concentration of N$^{+}$ poly-silicon layer, respectively.
The results suggest that the work function of TCO has little impact
on the cell and that the doping concentration of N$^{+}$ poly-silicon
layer is at a magnitude of 10$^{19}$19 cm$^{-3}$ and that the thickness
of N$^{+}$ polysilicon layer can vary from 20 \textgreek{m}m to 300
\textgreek{m}m. By simulating the P$^{+}$ $\alpha$-Si /N$^{+}$
polycrystalline solar cell we conclude that this is a prospective
solar cell with a high efficiency of 9.44\%, high fill factor of 0.845
and high Voc of 0.752 V if the optimization parameters are set. Therefore,
our study on the device formed by the P$^{+}$ $\alpha$-Si/N$^{+}$
poly-silicon is very significant in exploring the high efficiency
poly-silicon solar cell.

\section*{Acknowledgement}

The authors greatful acknowledge Professer S. Fonash of the Pennsylvania
State University and the Electric Power Research Institute for providing
the AMPS-1D program used in the simulations. This work was supported
by the Natural Science Foundation of Fujian Province of China (Grant
No. A0220001), Science Research Project of Leshan Vocational \& Technical
College (Grant No. KY2011001) and the Key Research Project in Science
and Technology of Leshan (Grant No. 2011GZD050).

\bibliographystyle{unsrt}
\bibliography{cite2}

\end{document}